\documentclass[12pt,preprint]{aastex}
\begin{document}

\title{A fundamental plane for {\bf long} gamma-ray bursts with X-ray plateaus}

\author{Dainotti, M. G. \altaffilmark{1,2,3}, Postnikov, S. \altaffilmark{4}, Hernandez, X., \altaffilmark{5}, Ostrowski, M. \altaffilmark{2}}
 
\altaffiltext{1}{Department of Physics \& Astronomy, Stanford University, Via Pueblo Mall 382, Stanford CA 94305-4060, E-mail: mdainott@stanford.edu}

\altaffiltext{2}{Obserwatorium Astronomiczne, Uniwersytet Jagiello\'nski, ul. Orla 171, 31-501 Krak{\'o}w, Poland, E-mails: dainotti@oa.uj.edu.pl,mio@oa.uj.edu.pl}

\altaffiltext{3}{INAF-Istituto di Astrofisica Spaziale e Fisica cosmica, c/o CNR - Area della Ricerca di Bologna, Via Gobetti 101, 40129 - Bologna, Italy, E-mails: mariagiovannadainotti@yahoo.it}

\altaffiltext{4}{The Center for Exploration of Energy and Matter, Indiana University, Bloomington IN 47405, USA: E-mail: postsergey@gmail.com}
\altaffiltext{5}{Instituto de Astronom\'{\i}a, Universidad Nacional Aut\'onoma de M\'exico, Ciudad de M\'exico 04510, M\'exico E-mail: xavier@astro.unam.mx}

\date{\today}
\begin{abstract}
A class of long Gamma-Ray Bursts (GRBs) presenting light curves with an extended plateau phase in their X-ray afterglows obeys a correlation between the rest frame end time of the plateau, $T_a$, and its corresponding X-ray luminosity, $L_{a}$, Dainotti et al. (2008). In this work we perform an analysis of a total sample of 176 {\it Swift} GRBs with known redshifts, exhibiting afterglow plateaus. By adding a third parameter, that is the peak luminosity in the prompt emission, $L_{peak}$, we discover the existence of a new three parameter correlation. 
The scatter of data about this plane becomes smaller when a class-specific GRB sample is defined. This sample of 122 GRBs is selected from the total sample by excluding GRBs with associated Supernovae (SNe), X-ray flashes and short GRBs with extended emission. {\bf With this sample the three parameter correlation identifies a GRB `fundamental plane'}. Moreover, we further limit our analysis to GRBs with lightcurves having good data coverage and almost flat plateaus, 40 GRBs forming our `gold sample'. The intrinsic scatter, $\sigma_{int}=0.27 \pm 0.04$, for the three-parameter correlation for this last subclass is more than twice smaller than the value for the $L_{a}-T_a$ one, making this the tightest three parameter correlation involving the afterglow plateau phase. Finally, we also show that a slightly less tight correlation is present between $L_{peak}$ and a proxy for the total energy emitted during the plateau phase, $L_a T_a$, confirming the existence of an energy scaling between prompt and afterglow phases.
\end{abstract}

 
\maketitle
\section{Introduction}
GRBs, very energetic events with typical isotropic prompt emission energies, $E_{iso}$ (erg), in the $10^{53} erg$ range, have been detected out to redshifts, z, of $\sim 10$ \citep{Cucchiara2011}. This last feature raises the tantalizing possibility of extending direct cosmological studies far beyond the redshift range covered by SNe. However, GRBs are not standard candles in any trivial way. Indeed, the number of sub-classes into which they are grouped has grown. GRBs are classified depending on their duration into short ($T_{90} \leq 2$ s) and long ($T_{90} \ge 2$ s)\footnote{where $T_{90}$ is the time interval over which between $5\%$ and $95\%$ of the total prompt energy is emitted.} \citep{Kouveliotou}. Later, a class of GRBs with mixed properties, such as short GRBs with extended emission (ShortEE), was discovered \citep{nb06}. Long GRBs, depending on their {\bf fluence (erg $cm^{-2}$)}, can be divided into normal GRBs or X-ray Flashes (XRFs), {\bf the latter are empirically defined as GRBs with a greater fluence in the X-ray band ($2-30$ keV) than in the $\gamma$-ray band ($30-400$ keV)}. In addition, several GRBs also present associated SNe, hereafter GRB-SNe. Regarding instead lightcurve morphology, a complex trend in the afterglow has been observed with the Swift Satellite \citep{Gehrels04,OB06} showing a flat part, the plateau, soon after the steep decay of the prompt emission. Along with these categories several physical mechanisms for producing GRBs have also been proposed. For example, the plateau emission has been mainly ascribed to millisecond newborn spinning neutron stars, e.g. Zhang \& M\'{e}sz\'{a}ros (2001), Troja et al. 2007, Dall'Osso et al. 2011, Rowlinson et al. (2013,2014), Rea et al. (2015) or to accretion onto a black hole (Cannizzo \& Gerhels 2009, Cannizzo et al. 2011). A promising field has been the search for correlations between relevant GRB parameters, e.g. Amati et al. (2002), Yonetoku et al. (2004), Ghirlanda et al. (2004), Ghisellini et al. (2008), Oates et al. (2009, 2012), Qi et al. 2009, Willingale et al. (2010), to attempt their use as cosmological indicators and to glimpse insights into their nature.

The correlations thus far discovered suffer from having large scatters \citep{Collazzi2008}, beyond observational uncertainties, highlighting that the events studied {\bf probably} come from different classes of systems {\bf or perhaps from the same class of objects, but we do not yet observe a sufficiently large number of parameters to characterize the scatter.} As the categories of GRBs have grown over the years, many with observed X-ray afterglows and measured redshift, the possibility of isolating single classes has appeared. This allows to derive tighter correlations, thus increasing the accuracy with which cosmological parameters can be inferred (e.g Cardone et al. (2009, 2010), Dainotti et al. 2013b, Postnikov et al. 2014), and yielding more stringent constraints on physical models describing them.

One of the first attempts to standardize GRBs in the afterglow parameters was presented in Dainotti et al. (2008,2010) where an approximately inversely proportional correlation between the rest frame end time of the plateau phase, $T_{a}$ (in previous papers $T^{*}_a$), and its corresponding luminosity $L_{a}$ was discovered. Dainotti et al. (2013a) proved through the robust statistical Efron \& Petrosian (1992) method, hereafter EP, that this correlation is intrinsic, and not an artefact of selection effects or due to instrumental threshold truncation, as it is also the case for the $L_{peak}-L_{a}$ correlation, (Dainotti et al. 2011b,2015b), where $L_{peak}$ is the peak luminosity in the prompt emission.

In this letter we show how a careful discrimination of plateau phase GRBs can be performed to isolate, using X-ray afterglow light curve morphology, a sub class of events which define a very tight plane in a three dimensional space of $(\log L_{a}, \log T_{a}, \log L_{peak})$. A three parameter correlation emerges with an intrinsic scatter, $\sigma_{int}$, of $24 \%$ less than the $L_a-T_a$ correlation for the sample of 122 long GRBs. When we choose a subsample of high quality data (40 GRBs, hereafter the gold sample), a further $38\%$ reduction in $\sigma_{int}$ appears. {\bf We also show through bootstrapping that the reduction in scatter is not an artefact of observational biases. Actually, Dainotti et al. (2010,2011a) have already demonstrated through a careful data analysis and Monte Carlo simulations that to reduce the scatter of this correlation an appropriate selection criterion related to observational GRB properties is more important than simply increasing sample size.} The $\sigma_{int}$ of the $L_a-T_a-L_{peak}$ correlation for the gold sample is $54\%$ smaller than the scatter of the $L_a-T_a$ correlation for the sample of 122 long GRBs. A slightly more scattered $L_{peak}, (L_{a}T_{a})$ correlation is also present, which together with an almost constant total energy within the plateau phase for the gold sample is indicative of a strong energy coupling between prompt emission and X-ray afterglow phase.
In section \S \ref{sample selection} and \S \ref{3D correlation} we describe the {\it Swift} data sample used and the three parameter correlation respectively. In section \S \ref{discussion} we present the $(L_{a} T_{a}, L_{peak})$ correlation as the tightest currently available involving the afterglow phase, together with our concluding remarks.

\section{Sample selection}\label{sample selection}

We analysed 176 GRB X-ray plateau afterglows detected by {\it Swift} from January 2005 up to July 2014 with known redshifts, spectroscopic or photometric, available in Xiao \& Schaefer (2009), in the Greiner web page \footnote{http://www.mpe.mpg.de/~jcg/grbgen.html} and in the Circulars Notice  (GCN), excluding redshifts for which there is only a lower or an upper limit. The redshift range of our sample is $(0.033, 9.4)$. We include all GRBs for which the Burst Alert Telescope (BAT)+ X-Ray Telescope (XRT) light curves can be fitted by the Willingale et al. (2007), phenomenological model, hereafter W07.
The W07 functional form for $f(t)$ is:

\begin{equation}
f(t) = \left \{
\begin{array}{ll}
\displaystyle{F_i \exp{\left ( \alpha_i \left( 1 - \frac{t}{T_i} \right) \right )} \exp{\left (
- \frac{t_i}{t} \right )}} & {\rm for} \ \ t < T_i \\
~ & ~ \\
\displaystyle{F_i \left ( \frac{t}{T_i} \right )^{-\alpha_i}
\exp{\left ( - \frac{t_i}{t} \right )}} & {\rm for} \ \ t \ge T_i \\
\end{array}
\right .
\label{eq: fc}
\end{equation}

\noindent for both the prompt (index `i=\textit{p}') $\gamma$\,-\,ray and initial X -ray decay and for the 
afterglow (``i=\textit{a}"), modeled so that the complete lightcurve $f_{tot}(t) = f_p(t) + f_a(t)$ 
contains two sets of four free parameters $(T_{i},F_{i},\alpha_i,t_i)$. The transition from the exponential 
to the power law occurs at the point $(T_{i},F_{i}e^{-t_i/T_i})$ where the two functional forms have 
the same value. The parameter $\alpha_{i}$ is the temporal power law decay index and the 
time $t_{i}$ is the initial rise timescale. We exclude the cases when the fitting procedure fails or when the determination of $1 \sigma$ confidence intervals does not fulfill the Avni (1976) $\chi^{2}$ prescriptions, see the xspec manual  \footnote{http://heasarc.nasa.gov/xanadu/xspec/manual/XspecSpectralFitting.html.}.
 We compute the source rest frame isotropic luminosity $L_a$ in units of $erg/s$ in the {\it Swift} XRT band pass, 
$(E_{min}, E_{max})=(0.3,10)$ keV as follows:

\begin{equation}
L_a= 4 \pi D_L^2(z) \, F_X (E_{min},E_{max},T_a) \cdot \textit{K} ,
\label{eq: lx}
\end{equation}
where $D_L(z)$ is the luminosity distance for the redshift $z$, assuming a flat $\Lambda$CDM cosmological model with $\Omega_M = 0.3$ and $H_0 = 70$ $km s^{-1} Mpc^{-1}$, $F_X$ is the measured X-ray energy flux in 
(${\rm erg/cm^{-2} s^{-1}}$) and  \textit{K} is the \textit{K}-correction for cosmic expansion $(1+z)^{(\alpha-1)}$.
The lightcurves are taken from the Swift web page repository, $http://www.swift.ac.uk/burstanalyser$ 
and we followed Evans et al. (2009) for the evaluation of the spectral parameters.
As shown in Dainotti et al. (2010) requiring an {\bf observationally homogeneous sample in terms of $T_{90}$ and spectral lag properties} implies removing short GRBs ($T_{90} \leq 2$ s) and ShortEE from the analysis. {\bf We remove the GRBs catalogued as ShortEE in Norris \& Bonnel (2006), Levan et al. (2007), Norris et al. (2010). For the removal of the remaining ShortEE GRBs we follow the definition of Norris et al. (2010), who identify ShortEE events as those presenting short spikes followed, within $10$ s, by  a drop in the intensity emission by a factor of $10^{3}$ to $10^{2}$, but with almost negligible spectral lag. Additionally, since there are long GRBs for which no SNe has not been detected, for example the nearby $z=0.09$ SNe-less GRB 060505, the existence of a new group of long GRBs without supernova has been suggested, thus highlighting the possibility of two types of Long-GRBs, with and without SNe.} Therefore, in the interest of selecting an {\bf observational} homogeneous class of objects, we consider only the long GRBs with no associated SNe. {\bf In this specific criteria all the GRB-SNe which follow the Hjorth \& Bloom (2011) classification are removed}. Similarly, {\bf to keep the sample homogeneous regarding the ratio between $\gamma$ and X-ray fluence, we removed all XRFs. The selection criteria are applied in the observer frame.}
Figure 1 shows the light curve for GRB 061121 with the best fit model light curve superimposed. The plateau phase is clearly seen between $2.6$ and $3.8$ in $\log(T)$ in units of seconds (s).


\begin{figure}
\includegraphics[width=0.9\hsize,height=0.8\textwidth,angle=0,clip]{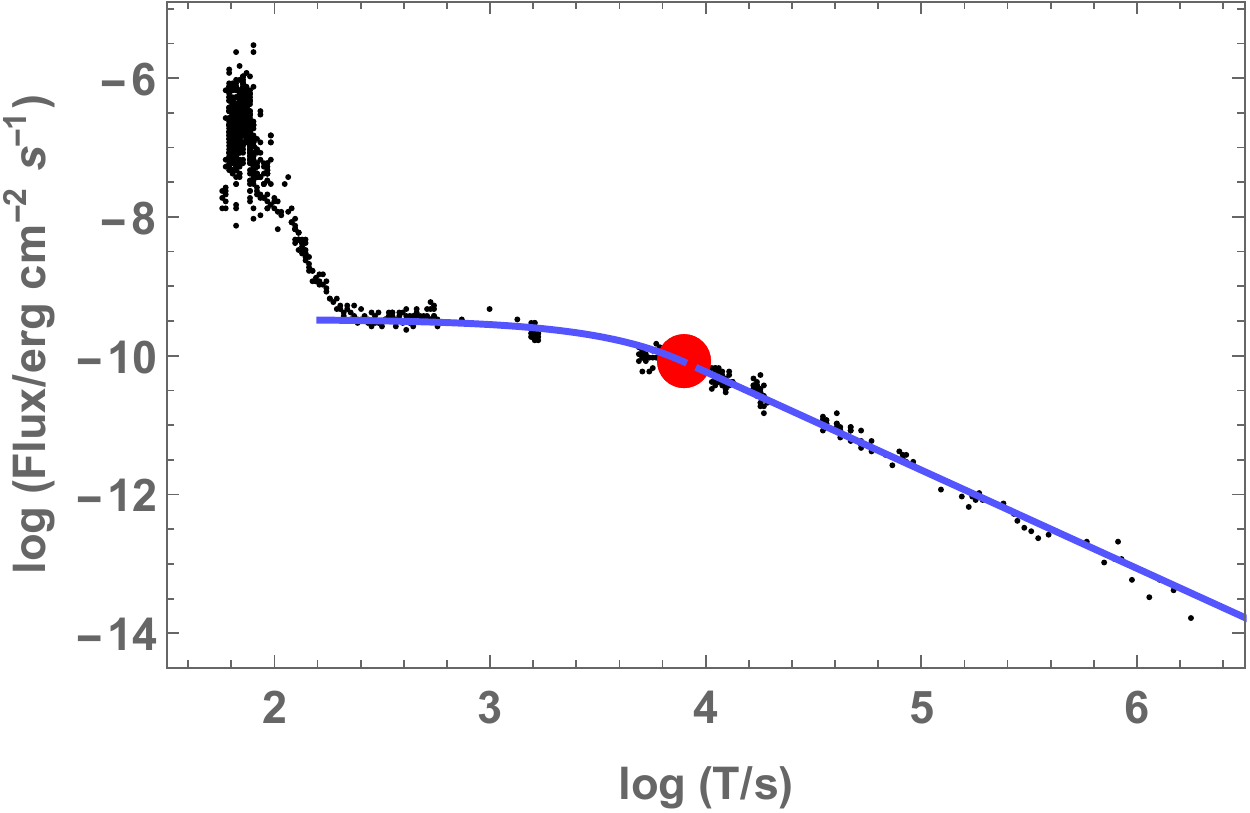}
\caption{The observed light curve for GRB 061121 with the best fit W07 model superimposed. The red dot marks the end of the plateau phase in the X-ray afterglow. \label{fig1}}
\end{figure}

In all that follows $L_{peak}$($erg/s$) is defined as the prompt emission peak flux over a $1$ s interval. Following Schaefer (2007) we compute $L_{peak}$ as follows:

\begin{equation}
L_{peak}= 4 \pi D_L^2(z) \, F_{peak} (E_{min},E_{max},T_a) \cdot \textit{K} ,
\label{eq: lx}
\end{equation}
where $F_{peak}$ is the measured gamma-ray energy flux over a $1$ s interval (erg $cm^{-2} s^{-1}$).
To make the sample for this analysis more homogeneous regarding the spectral features, we consider only the GRBs for which the spectrum computed at $1$ second has a smaller $\chi^2$ for a single power law (PL) fit than for a cutoff power law (CPL). Specifically, following Sakamoto et al. (2010) when the $\chi_{CPL}^{2}-\chi_{PL}^{2}<6$, the PL fit is preferred. We additionally discard $6$ GRBs which were better fitted with a Black Body model than with a PL. 
This full set of criteria reduces the sample to $122$ long GRBs.
Finally, we construct a sub sample by including strict data quality and morphology criteria: at least $5$ points should be at the beginning of the plateau and {\bf the steep plateaus (the angle of the plateau greater than $41$ degrees), which constitute the $11\%$ of the total sample and are the high angle tail of the distribution, are removed}. {\bf The first of the above selection criteria guarantees that the lightcurves clearly present the transition from the steep decay after the prompt to the plateau. The number of points required for the W07 fit should be at least $4$, since there are $4$ free parameters in the model, one of which should be after the end of the plateau. Thus, the requirement of $6$ points in total ($5$ at the start and at least one after the plateau) ensures a minimum number of points to have both a clear transition to the plateau phase (in fact, in some cases $3$ points do not offer a wide enough time range to determine the start of the plateau) and simultaneously to constrain the plateau.} This data quality cut defines the gold sample, which includes $40$ GRBs. We have also checked through the T-test that this gold sample is not statistically different from the distribution of ($L_a$, $T_a$, $L_{peak}$, $z$) of the full sample, thus showing that the choice of this sample does not introduce any biases, such as the Malmquist or Eddington ones, against high luminosity and/or high redshift GRBs. {\bf Specifically, $L_a$,$T_a$,$L_{peak}$ and $z$ of the gold sample present similar Gaussian distributions, but with smaller tails than the total sample (see Dainotti et al. 2015a), thus there is no shift of the distribution towards high luminosities, larger times or high redshift. So, the selection cut naturally removes the majority of the high error outliers of the variables involved, thus reducing the scatter of the correlation for the gold sample.} 

\section{The $(L_{a}, T_{a}, L_{peak})$ parameter space}\label{3D correlation}

The left panel of figure (2) shows 176 GRBs in the $(L_{a}, T_{a}, L_{peak})$ parameter space, where distinct sub-classes of GRBs show greater spread about the plane than the gold sample. The right panel in figure (2) shows the fundamental plane in projection for the 122 long GRBs, the reduction in the intrinsic scatter is clear. 

To explore if the two-dimensional $L_a-T_a$ correlation is the projection of a tighter $(L_{a}-T_{a}-L_{peak})$ plane, we plot the 122 long GRBs, in a $(L_{a}, T_a)$ plane, binned according to their $L_{peak}$ values into three equally populated ranges : $49.9 \leq \log L_{peak} \leq 51.4$, $51.4 \le \log L_{peak} \leq 51.8$ and $51.8 \le \log L_{peak} \leq 53.0$, red circles, blue squares and black triangles, respectively, in the left panel of figure (3). For reference, the curves show best fit lines of fixed slope equal to one and free intercept calculated for each $L_{peak}$ bin. We see a clear monotonic trend, in that the intercept of the lines is determined by the $L_{peak}$ bin of the sub-sample, all of which show a significantly smaller dispersion than the 122 GRB sample. {\bf The above is indicative of an underlying plane in the $(L_{a}, T_a, L_{peak})$ parameter space, the $L_{a}-T_a$ correlation being just a projection of it. Introducing a third (prompt emission) parameter, $L_{peak}$, reduces the $L_a-T_a$ correlation scatter, in part associated to the prompt luminosity.}

Parametrizing this plane using the angles $\theta$ and $\phi$ of its unit normal vector gives: 

\begin{equation}
\log L_a = C_o - \cos(\phi) \tan(\theta) \log T_a - \sin(\phi) \tan(\theta) \log L_{peak},
\end{equation}
\noindent where $C_o = C(\theta,\phi,\sigma_{int})+z_o$ is the normalization of the plane correlated with the other variables, $\theta$,$\phi$ and $\sigma_{int}$; while $z_o$ is the uncorrelated fitting parameter related to the normalization and $C$ is the covariance function. This normalization of the plane allows the resulting parameter set, $\theta$, $\phi$, $\sigma_{int}$ and $z_o$ to be uncorrelated and provides explicit error propagation.
Accounting for all the error propagation we fit an optimal plane for the gold sample distribution given by:
\begin{equation}
\log L_{a}=(15.75 \pm 5.3) - (0.77 \pm 0.1) \log T_{a} + (0.67 \pm 0.1)\log L_{peak},
\end{equation}

\noindent 
where $C_o=(15.75 \pm 5.3)$, $\cos(\phi) \tan(\theta)=-(0.77 \pm 0.1)$ and $\sin(\phi) \tan(\theta)=(0.67 \pm 0.1) \pm 0.1$. 
All the fits presented in the paper are performed using the D'Agostini method \citep{Dago05} with $1 \sigma$ uncertainties on the coefficients given; $\sigma_{int}=0.27 \pm 0.04$ is reduced by $36\%$ when compared to the $(L_{a} -T_a)$ correlation for the gold sample.
The adjusted $R_{adj}^2$ for the gold sample is $0.80$. $R_{adj}$ gives a modified version of the coefficient of determination, $R^2$, adjusting for the number of parameters in the model. $R^2=0.81$, the Pearson correlation coefficient, $r$, is $0.90$ with a probability of the same sample occurring by chance, $P=4.41 \times 10^{-15}$.
The normalization of the plane, $C(\sigma_{int}, \phi, \theta)$, is given by:
\begin{equation}
C = 18.30  - 59.90 \theta^2 - 0.29 \sigma_{int} +  0.27 \sigma_{int}^2 - 4.11 \phi - 0.06 \sigma_{int} \phi + 
 14.97 \phi^2 + \theta (92.07 - 0.09 \sigma_{int} +  84.85 \phi).
\end{equation}
\noindent  

\begin{figure}[!t]
\hskip -60pt
\includegraphics[width=0.65\hsize,height=0.35\textwidth,angle=0,clip]{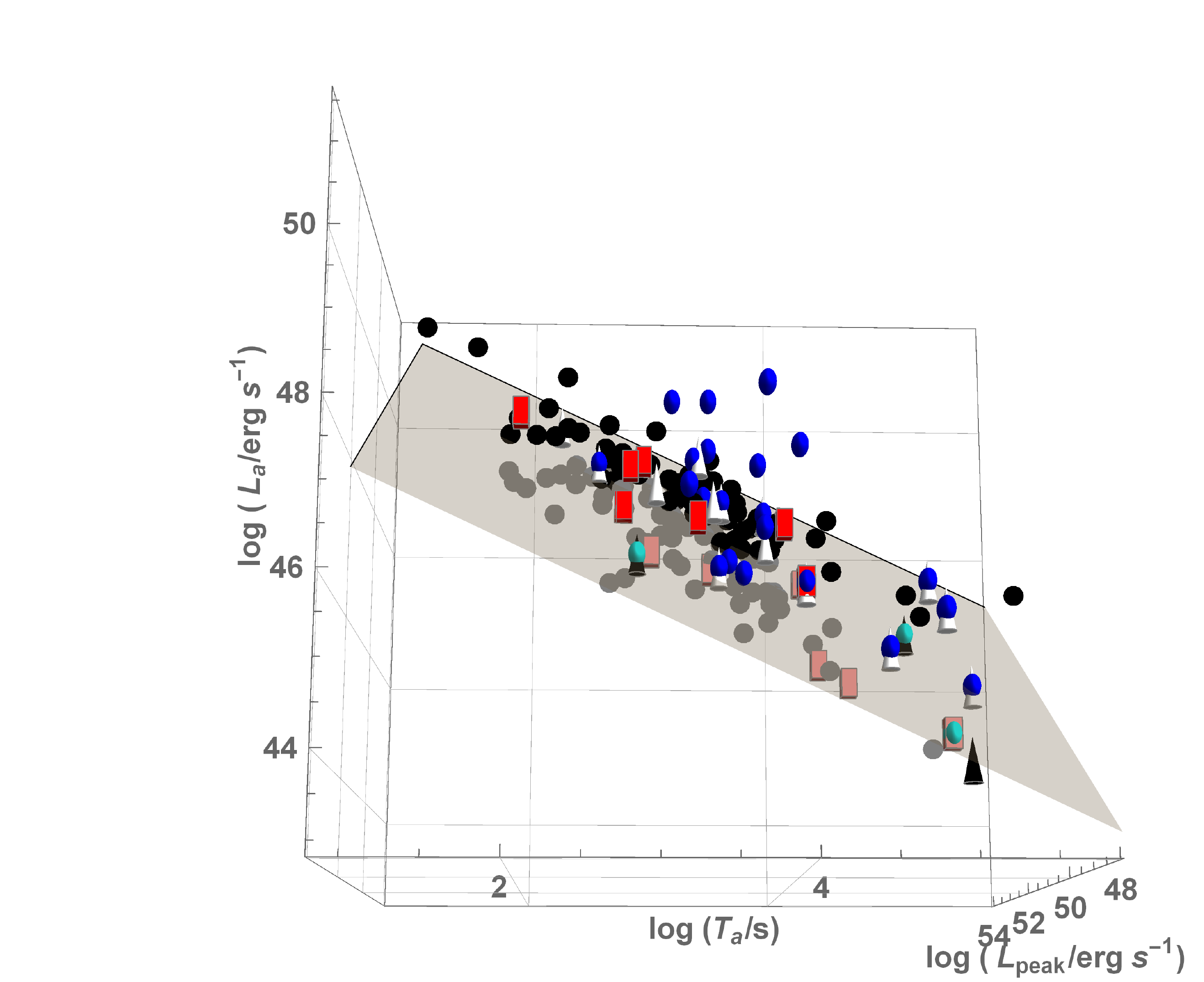}
\hskip -7pt
\includegraphics[width=0.60\hsize,height=0.40\textwidth,angle=0,clip]{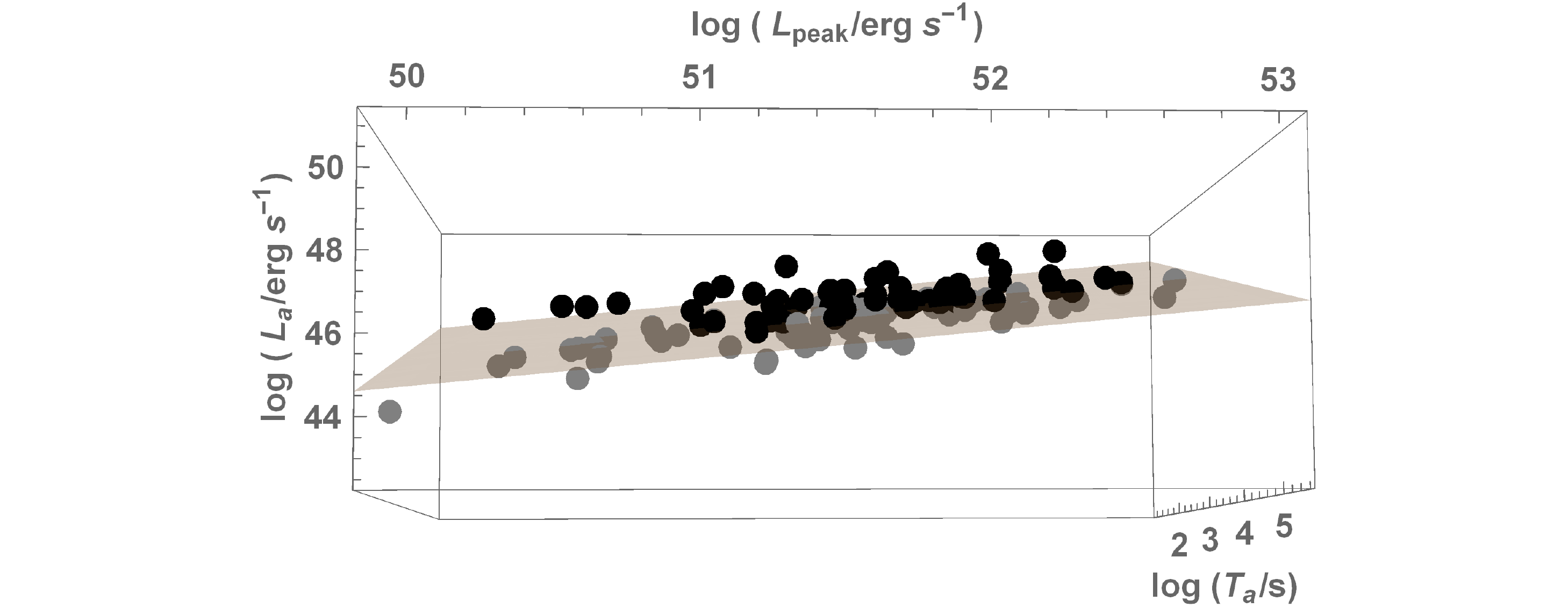}
\caption{The left panel shows 176 GRBs in the $(L_{a}, T_{a}, L_{peak})$ space, with the fitted plane including GRB-SNe (white cones), X-ray flashes (blue spheres), shortEE (red cuboid) and long GRBs (black circles). The right panel shows a much tighter plane which results by including only the $122$ long GRB sample. The gray and black circles are GRBs which lie below and above the plane respectively.}
\end{figure}

\noindent For the 122 GRBs, results are:
\begin{equation}
\log L_{a}=(15.69 \pm 3.8) + (0.67 \pm 0.07) \log L_{peak} - (0.80 \pm 0.07) \log T_{a},
\end{equation}
\noindent with $\sigma_{int}=0.44 \pm 0.03$. Thus, the reduction in $\sigma_{int}$ from the 3D correlation for the sample of 122 GRBs to the 3D correlation for the gold sample is again $36\%$. 
The $R_{adj}^2$ for this distribution is $0.56$, $R^2=0.57$, $r=0.93$ with $P \le 2.2 \times 10^{-16}$.
Finally, $\sigma_{int}$ of the 3D correlation for the gold sample is $54\%$ smaller than the one in the 2D correlation for the 122 GRB sample.

The plane can be visualized edge-on in an infinite number of projections, according to how the projection angle is rotated. We choose a projection where the plane is seen edge-on and one of the axes contains only one of the three relevant parameters, see the right panel of figure (3), which shows the plane for the gold sample. By comparing it with figure (2), and noting the change in scales, it is obvious that $\sigma_{int}$ has been substantially reduced, {\bf although a few outliers remain, keeping the scatter larger than measurement errors}.

\begin{figure}
\includegraphics[width=0.49\hsize,height=0.46\textwidth,angle=0,clip]{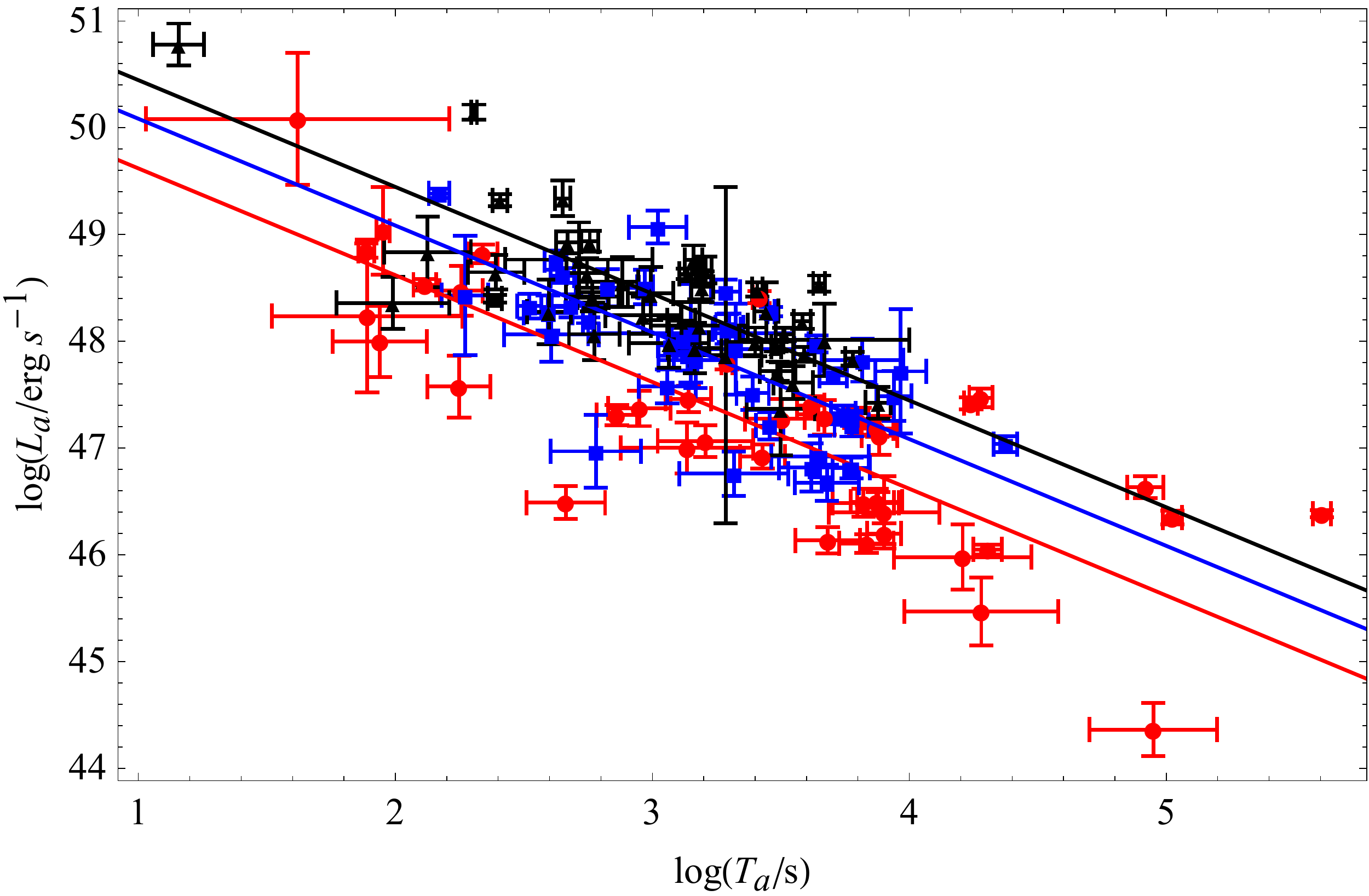}
\includegraphics[width=0.51\hsize,height=0.50\textwidth,angle=0,clip]{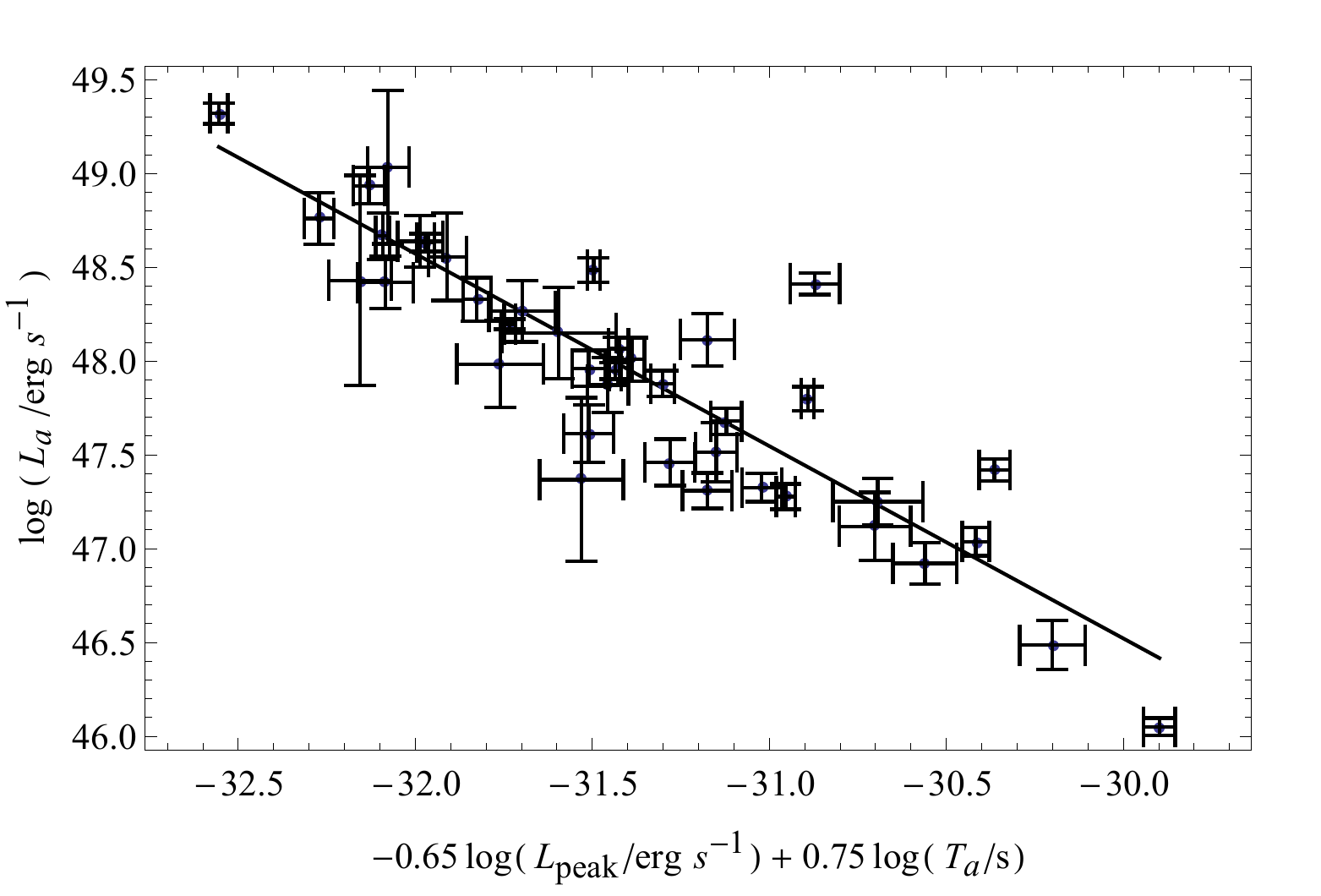}
\caption{Left panel: the ($L_{a}, T_{a}$) plane with error bars, binned into 3 equally populated $L_{peak}$ ranges; $49.9 \leq \log L_{peak} \leq 51.4$ (red circles), $51.4 \le \log L_{peak} \leq 51.8$ (blue squares) and $51.8 \le \log L_{peak} \leq 53.0$ (black triangles). The right panel shows the edge-on projection along the intrinsic plane for the ($L_a,T_a,L_{peak}$) correlation for the gold sample.}
\end{figure}

\begin{figure}
\plotone{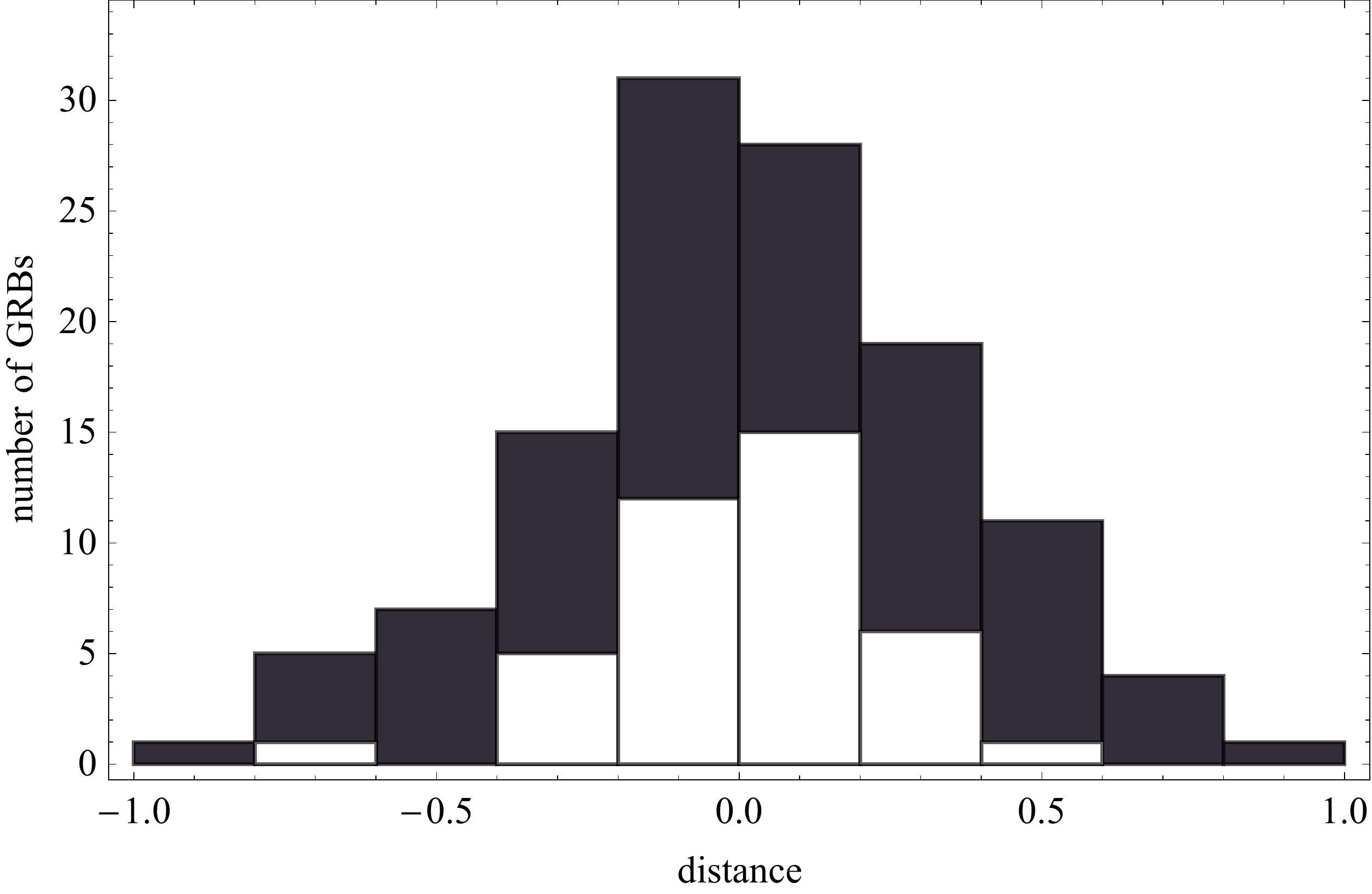}
\caption{Histogram for the 122 long GRBs with distances to the best fit plane with different shading for gold (white) and non-gold (black) GRBs.}
\end{figure}
To analyze the distributions of GRBs about the plane, we compute their geometric distance to it for the 122 GRBs and the gold sample, see
figure (4). The latter sample is less scattered about the plane than the first. This result is not due to the reduced sample size, as checked by $10^{4}$ Monte Carlo simulations using bootstrapping of 40 GRBs from the total sample. The probability of obtaining such a random sample having an intrinsic scatter $\sigma_{int} \leq 0.27$ is of $~0.3\%$.  {\bf Although we have considered all known biases, it cannot be ruled out completely that part of the reduction in scatter might be attributed to some unknown bias.}

\section{Discussion and comparison with other extended $L_a-T_a$ correlations}\label{discussion}

\begin{figure}
\plotone{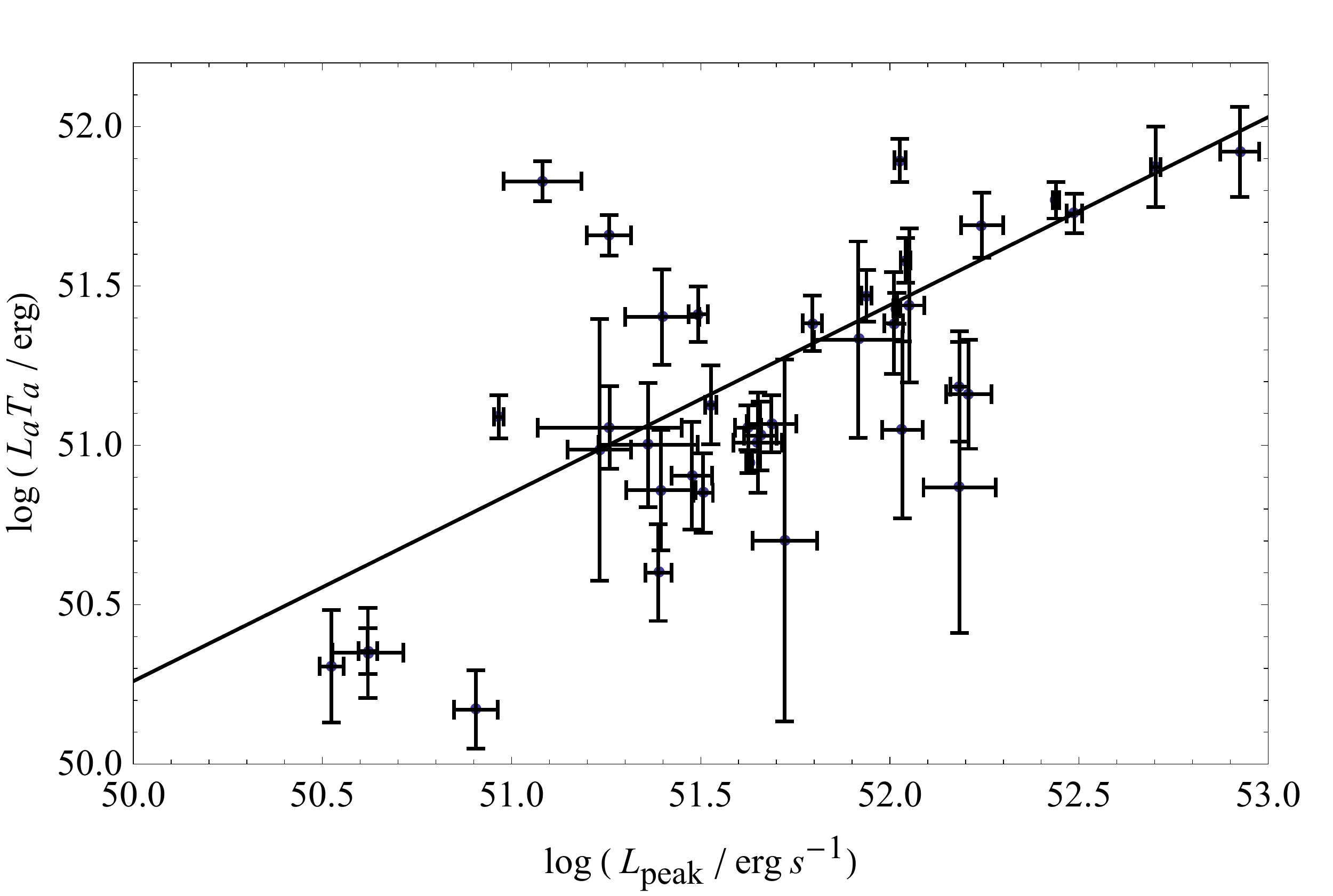}
\caption{($L_{a}T_{a}$,$L_{peak}$) distribution for the 40 GRBs in the gold GRB sample.}
\end{figure}

To derive insights into the physical nature of the link between prompt and afterglow parameters evident in the plane obtained, we explore the relation between a proxy of the plateau energy, $L_{a}T_a$ and $L_{peak}$. 
In figure (5) we show that $(L_{a}T_{a}) \propto L_{peak}^{0.59}$, with $r=0.60$ and $P=1.9 \times 10^{-13}$ for the gold sample, while $r=0.70$  and $P=4.2 \times 10^{-7}$ for the 122 GRBs. This result demonstrates that the prompt kinetic power is strongly correlated with the plateau energy, for the well-defined plateau exhibiting GRBs. The best fit slope with $\sigma_{int}=0.29$ is:

\begin{equation}
\log(L_{a}T_{a}) = 20.63 + 0.59 \log (L_{peak})
\end{equation}

This correlation is different from the one presented in Bernardini et al. (2012), where $L_{a}T_{a}$ and two additional parameters, $E_{peak}$ and $E_{iso}$ is explored yielding $\sigma_{int}=(0.31 \pm 0.03)$. 
Another plane ($L_{peak}-E_{iso}-E_{peak}$) has been defined, but only amongst prompt emission parameters (Tsutsui et al. 2009). 
We compare the $L_a-T_a-L_{peak}$ correlation with other three parameter correlations, which are extensions of the $L_a-T_a$ correlation.
Xu \& Huang (2011) obtained a tighter $(L_{a}-T_{a}-E_{iso})$ correlation with $\sigma_{int}=0.43$, as compared to the ($L_{a}-T_{a}$) one which yielded $\sigma_{int}=0.85$ for their sample. 
The $\sigma_{int}$ of our $(L_a,T_a,L_{peak})$ plane is smaller by $37\%$ than the $\sigma_{int}$ of the $(L_{a}-T_{a}-E_{iso})$ plane. 

In fact, Dainotti et al. (2011b, 2015b) showed that $L_{a}-L_{iso}$, where $L_{iso}= E_{iso}/T_{90}$, correlates better than 
$L_{a}-E_{iso}$ and that $L_{a}-L_{peak}$ correlates better than $L_{a}-L_{iso}$, respectively. Thus, $L_{peak}$ is more tightly related to $L_{a}$ than any other prompt $L_{iso}$. The above suggests including $L_{peak}$ and not $E_{iso}$ as a third parameter in the search for a three parameter correlation. 
Indeed, Dainotti et al. (2015b) have demonstrated through the EP method, that the $L_a-L_{peak}$ correlation is intrinsic and not due to any selection bias.
Thus, from the intrinsic nature of the $L_{peak}-L_a$ and the $L_a-T_a$ correlations, it follows that the ($L_a,T_a,L_{peak}$) correlation is also intrinsic. 

Another extension of the $L_a-T_a$ correlation ($T_a$,$L_a$,$E_{peak}$) is presented in Izzo et al. (2015) where $\sigma_{int}=0.34$. This scatter is larger by $21\%$ than the $\sigma_{int}=0.27$ for the ($L_a$,$T_a$,$L_{peak}$) plane. {\bf Notice that $L_{peak}$ is a more suitable variable than $E_{peak}$, as $L_{peak}$ is subject to only low luminosity truncation and leads to the intrinsic $L_{peak}-L_a$ correlation (Dainotti et al. 2015b). On the other hand, $E_{peak}$ can introduce biases due to threshold limits both at low and high energies, see  Lloyd \& Petrosian (1999), and its intrinsic distribution, which would possibly allow a bias-free ($T_a$,$L_a$,$E_{peak}$) correlation, has not yet been determined.}

To conclude, isolating 40 long GRBs (without associated SNe and excluding also XRFs) with well-defined plateaus we obtain a 3D correlation significantly tighter ($54\%$) than the 2D correlation for the 122 long GRBs. 
This correlation can be a useful tool to reduce the uncertainties in inferred cosmological parameters in the high redshift range accessible only to GRBs. {\bf Additionally, it can further constrain GRB physical models that connect prompt and afterglow plateau properties. It is also worth investigating if the $(L_{a}-T_{a}-E_{iso})$ and ($L_{a}-T_{a}-L_{peak}$) correlations might both be the reflection of the same underlying physics (Shao \& Dai 2007 and Wang et al. 2016).}

\section{Acknowledgments}
This work made use of data supplied by the UK Swift Science Data Centre at the University of Leicester. M.G.D acknowledges the Marie Curie Program, because the research leading to these results has received funding from the European Union Seventh Framework Program (FP7-2007/2013) under grant agreement N 626267.  M. O. acknowledges the Polish National Science Centre through the grant DEC-2012/04/A/ST9/00083. X. H. acknowledges UNAM-DGAPA IN100814 and CONACyT.

\end{document}